\documentclass[preprint,preprintnumbers]{revtex4}

\usepackage{graphicx}

\usepackage{amssymb}

\usepackage{mathrsfs}

\begin{document}

\title{Fractional Topological Insulators- A Bosonization Approach} 

\author{D. Schmeltzer}

\affiliation{Physics Department, City College of the City University of New York,  
New York, New York 10031, USA}

\begin{abstract}

A metallic disk with strong  spin  orbit interaction  is investigated  . The finite disk geometry introduces    a confining  potential.   Due to the strong spin-orbit interaction and confining potential the metal disk is described by an effective one dimensional  with a harmonic potential.  The harmonic potential gives rise to classical turning points. As a result open boundary conditions must be used. We Bosonize the model and obtain   chiral Bosons  for each spin on the edge of the disk. When the   filling fraction  is reduced  to  $\nu=\frac{k_{F}}{k_{so}}=\frac{1}{3}$ the electron- electron  interactions are studied using  the    Jordan Wigner  phase for  composite fermions which gives rise to a Luttinger liquid. When the metallic disk is in the proximity with a superconductor a Fractional Topological Insulators is obtained.

An experimental realization is proposed. We show that by tunning the chemical potential we control the classical turning points for which a  Fractional Topological Insulator is realized. 

\vspace{0.3 in}  


\end{abstract}

\maketitle

\vspace{0.2 in}
\noindent
\textbf{1. Introduction}
 
\vspace{0.2 in}
\noindent
The  presence of the spin-orbit interaction in confined geometries  gives rise to a Topological Insulator ($T.I.$).  Following  ref. \cite{Stern} one maps   the spin-orbit interaction  to a  spin dependent magnetic field $B\sigma_{z}$ . As a result the non interacting electrons  are mapped to two effective  Quantum Hall  problems,  for each species of spin. When the electron density  is tuned to an integer Landau  filling   $\nu=k$ (for each spin) the ground state is  made up of two decoupled spin species which   forms an  integer Quantum Hall states with opposite chiralities . When $k$ is odd the system is a $TI$  and when $k$ is even we have a trivial insulator. When $\nu=\frac{1}{k}$  the presence of electron-electron interaction for each species of spin gives rise to a Fractional Topological Insulators ($F.T.I.$).

\noindent 
Using   the  proposal for the Fractional  Quantum Hall, built  on  an array of quantum wires \cite{Kane} the authors 
  \cite{Klinovaja,Halperin} have shown that by fine tuning the spin orbit interaction for  a configuration of coupled chains   a  Topological Insulator ($T.I.$)   emerges. When  the  filling factor is   such  that it corresponds to     composite Fermions, a Fractional Topological Insulator   ($F.T.I.$)  has been introduced   in ref. \cite{Halperin}. It has been shown that for a model of coupled  chains  in the $y$ direction the    spin orbit interaction can be gauged away resulting in   twisted  boundary conditions for which a $F.T.I.$ was obtained.

The purpose of this paper is two demonstate that for 
a  two dimensional  metallic disk with spin-orbit interaction and electron electron interaction gives  rise either to a Topological Insulator or Fractional Topological  when the disk is in the proximity to  a superconductor.
We Bosonize \cite{Haldane,Shankar,Luther}the model in the limit  of  strong spin orbit   interactions and geometrical confinement . We find that  the edge of  the disk   is equivalent to a one dimensional model with a harmonic potential.
We  obtain a chiral Bosonic model  \cite {Jackiv,Jacobs} and show that a T.I. emerges for the filling factor  $\nu=\frac{k_{F}}{k_{so}}=1$.

\noindent
For the filling factor  $\nu=\frac{k_{F}}{k_{so}}=\frac{1}{3}$ we  use   the  composite electrons  method \cite{Jain,Basu}   and  show that the  composite Jordan Wigner phase \cite{david}   gives rise  rise to an interacting one dimensional model  in the Bosonic  form  we obtain a Luttinger liquid with the Luttinger parameter $\kappa=\nu=\frac{k_{F}}{k_{so}}=\frac{1}{3}$ is which In the proximity to  a superconductor we obtain   a  $ F.T.I.$

\noindent
An experimental verification is proposed.
We show that the F.T.I. is obtained by tunning the chemical potential, the interactions and the radius of the disk.

The plan of the paper is as follows. In section $2$ we present the spin-orbit interactions in metals. In section  $2.1$ and $2.2$ we  review  the model   introduced in ref.\cite{Halperin} . We  find it advantageus to use  open boundary conditions and study the model  in the framework of  Bosonization.
In section $3.1$ we introduce  our new  model.  We  consider a metallic disk with srong  spin orbit interaction and confinement. In section $3.2$ we study the   
metallic disk with strong  spin orbit interactions and confinement for the  filling factor $\nu=\frac{k_{F}}{k_{so}}=\frac{1}{3}$  Using the composite Fermion  method  we obtain a Luttinger liquid which in the proximity with a superconductor  represents  a  $ F.T.I.$.
At the end of this section we consider the experimental realization of the model.
Section $4$ is devoted to conclusions.

\vspace{0.2 in}
\noindent
\textbf{2. The spin orbit in two dimensions in the  presence of a confining potential}

\vspace{0.2 in}
\noindent
The Hamiltonian for a two dimensional metal in the presence of a parabolic confining potential  is given by:
\begin{equation}
H=\frac{1}{2m^*}\Big[\vec{p}-\frac{\hat{\mu}}{2c}(\vec{\sigma}\times\vec{E})\Big]^{2}-\mu+V(\vec{x},\vec{y}) 
\label{spin orbit}
\end{equation}
Using the confining potential $ V(\vec{x},\vec{y})=\frac{\gamma}{2}(x^2+y^2)$  we obtain the electric field : $E_{x}=\gamma x$ , $E_{y}=\gamma y$  and $E_{z}=0$.
We introduce a  $fictitious$ magnetic field  $\frac{k_{so}}{a}\equiv\frac{\mu}{2 \hbar c}\equiv {B}$. As a result the Hamiltonian in Eq.$(1)$ is a function  of the spin orbit momentum $ k_{so}$  and takes the form: 
\begin{equation}
H=\frac{\hbar^{2}}{2m^*}\Big[(-i\partial_{x}-\sigma_{z}k_{so} \frac{ y}{a})^2+ (-i\partial_{y}+\sigma_{z}k_{so} \frac{ x}{a})^2\Big]-\mu+V(\vec{x},\vec{y})
\label{spinfictious}
\end{equation}
 $a$ is the lattice constant and    $A_{x}=\frac{k_{so}}{a}y$ , $A_{y}=\frac{k_{so}}{a}x$ are  the  gauge fields.

\vspace{0.1 in}
\noindent
\textbf{2.1.The emerging topological insulator for a  two dimensional  model  periodic in the $y$   direction   with the filling factor  $\nu=\frac{k_{F}}{k_{so}}=1$  for a system of coupled chains}

\vspace{0.1 in}
In this section we will review the model  introduced in ref. \cite{Halperin}.
We find essential to modify  the model   and use   open boundary conditions.
This modification is important for avoiding  complications caused by the twist  introduced by the spin-orbit interaction. The open boundary conditions impose a constraint on the Bosonic fields (the right and left Bosonic field are not independent).

For the remaining part we  will Bosonize  \cite{Jackiw} the model given in ref.  \cite{Halperin} using the open boundary conditions. We will use open boundary conditions also for the metalic disk (see sections $3.1-3.2$)
 The methodology for  both  model  will be same, therefore we find it necessary  to present the  details of the  Bosonization method (for open boundary conditions).
\noindent 
The model considered  in ref.  \cite{Halperin} is as follows:
 In the $y$ direction we have   $N$ chains  with the tunneling matrix element $t$.   The confining  potential $V(x,y)$  obeys,  for $0< x<L$  $V(x,y)=0$  and  for $| x|>L$   $V(x,y)\rightarrow \infty $.    We will assume open boundary conditions in the $x$ direction.  In the $y$ direction  the confined potential is effectively zero for $ 0<y<Na$ and  $V(x,y)\rightarrow \infty$   for   $y>Na$ ( $N$  are the number of  chains).  We will use  the conditions  $A_{x}=\frac{k_{so}}{a}y$ and $A_{y}=0$.
\begin{eqnarray}
&& H=H_{0}+H_{t} \nonumber\\&&
H_{0}=\sum_{\sigma=\uparrow,\downarrow}\sum_{n=1}^{N}\int\,dx \Psi^{\dagger}_{n,\sigma}(x)\Big[\frac{\gamma}{2}(-i\partial_{x}-\sigma_{z}k_{so}\frac{ y}{a})^2 -\mu +V(x)\Big]\Psi_{n,\sigma}(x),  \hspace{0.01 in} \frac{\hbar^{2}}{2m^*}= \frac{\gamma}{2}\nonumber\\&&
H_{t}=-t\sum_{\sigma=\uparrow,\downarrow}\sum_{n=1}^{N}\int\,dx \Psi^{\dagger}_{n,\sigma}(x)\Psi_{n+1,\sigma}(x)+h.c.\nonumber\\&&
\end{eqnarray}
$H_{0}$ is the one dimensional model for each chain and $H_{t}$ describes the tunneling between the chains.
The confining potential $V(x,y)$ enforces  the  open boundary conditions.
\begin{equation}
 \Psi_{n,\sigma}(x=0)= \Psi_{n,\sigma}(x=L)=0,\hspace{0.1 in} n=1,2...N
\label{eqa}
\end{equation}
The open  boundary conditions  avoid the   twist.  
\begin{eqnarray}
&&\Psi_{n,\sigma}(x)=e^{i \sigma_{z}k_{so}\frac{y}{a}x}\tilde{\Psi}_{n,\sigma}(x);\hspace{0.05 in} \Psi^{\dagger}_{n,\sigma}(x)=e^{-i \sigma_{z}k_{so}\frac{y}{a}x}\tilde{\Psi}^{\dagger}_{n,\sigma}(x)\nonumber\\&&
\tilde{\Psi}_{n,\sigma}(x=0)= \tilde{\Psi}_{n,\sigma}(x=L)=0,\hspace{0.1 in} n=1,2...N\nonumber\\&&
\end{eqnarray}
As a result  the Hamiltonian is transformed,
\begin{eqnarray}
&& H=H_{0}+H_{t}  \nonumber\\&& 
H_{0}=\sum_{\sigma=\uparrow,\downarrow}\sum_{n=1}^{N}\int\,dx \Psi^{\dagger}_{n,\sigma}(x)\Big[\frac{\gamma}{2}(-i\partial_{x}-\sigma_{z}k_{so}\frac{y}{a})^2 -\mu\Big]\Psi_{n,\sigma}(x)=\nonumber\\&&
\sum_{\sigma=\uparrow,\downarrow}\sum_{n=1}^{N}\int\,dx\tilde{\Psi}^{\dagger}_{n,\sigma}(x)\Big[\frac{\gamma}{2}(-i\partial_{x})^2 -\mu\Big]\tilde{\Psi}_{n,\sigma}(x)\nonumber\\&&
H_{t}=-t\sum_{\sigma=\uparrow,\downarrow}\sum_{n=1}^{N}\int\,dx \Psi^{\dagger}_{n,\sigma}(x)\Psi_{n+1,\sigma}(x)+h.c.=\nonumber\\&&
-t\sum_{\sigma=\uparrow,\downarrow}\sum_{n=1}^{N}\int\,dx \tilde{\Psi}^{\dagger}_{n,\sigma}(x) e^{i \sigma_{z}k_{so}\frac{(y(n)-y(n+1))}{a}x}\tilde{\Psi}_{n+1,\sigma}(x)+h.c.\nonumber
\end{eqnarray}
Next we use the mapping   $ \frac{ y(n)}{a}=2n-1$ for $ n=1,2...N$  introduced by  \cite{Oreg,Halperin}. This parametrization removes the oscillating  phase for certain channels  $n,n+1$ and  therefore gaps are opened. 

\noindent
In the next step we Bosonize  the model for  the filling factor $\nu=\frac{k_{F}}{k_{so}}=1$ using  the electronic density $n_{n,\sigma}(x)$  and the Bosonic phase $\varphi_{n,\sigma}(x)$ for each chain   ( $\mu$  is the chemical potential,$k_{F}$ is  the Fermi  momentum  and $k_{so}$ is the spin orbit  strength).
\begin{eqnarray}
&&\tilde{\Psi}_{n,\sigma}(x)=\frac{1}{\sqrt{2\pi a}}F_{n,\sigma}e^{\pm i\pi \int_{-\infty}^{x}\,dx' n_{n,\sigma}(x')}e^{-i\sqrt{\pi}\varphi_{n,\sigma}(x)},\hspace{0.01 in} n_{n,\sigma}(x)=\bar{n}+\frac{1}{\sqrt{\pi}}\partial_{x}\theta_{n,\sigma}(x)\nonumber\\&&
 \pi\bar{n}_{n,\sigma}=k_{F}, \partial_{x}\theta_{n,\sigma}(x)\equiv p_{n}(x),[\theta_{n,\sigma}(x),p_{n,'\sigma'}(x')]_{-}=i\delta(x-x')\delta_{n,n'}\delta_{\sigma,\sigma'}\nonumber\\&&
\tilde{R}_{n,\sigma}(x)=\frac{1}{\sqrt{2\pi a}}F_{n,\sigma}e^{i\sqrt{\pi}(\theta_{n,\sigma}(x)-\varphi_{n,\sigma}(x))}\equiv \frac{1}{\sqrt{2\pi a}} F_{n,\sigma} e^{i\sqrt{4\pi}\theta^{R}_{n,\sigma}(x)},\nonumber\\&&
\tilde{L}_{n,\sigma}(x)=\frac{1}{\sqrt{2\pi a}}F_{n,\sigma}e^{-i\pi(\theta_{n,\sigma}(x)+\varphi_{n,\sigma}(x))}\equiv  \frac{1}{\sqrt{2\pi a}}F_{n,\sigma} e^{-i\sqrt{4\pi}\theta^{L}_{n,\sigma}(x)}\nonumber\\&&
\tilde{\Psi}_{n,\sigma}(x)=e^{ik_{F}x}\tilde{R}_{n,\sigma}(x)+e^{-ik_{F}x} \tilde{L}_{n,\sigma}(x)\nonumber\\&&
\end{eqnarray}
$F_{n,\sigma}$, $F^{\dagger}_{n,\sigma}$ are anti commuting Klein factors \cite{Neupert,david,Stone}.

\noindent
Due to the boundary conditions  the left and right movers are not independent.
The Bosonic representation of the Fermion field  $ \tilde{\Psi}_{n,\sigma}(x)$ is given in terms of the right movers $\tilde{R}_{n,\sigma}(x)$ . The left movers are given by  $ \tilde{L}_{n,\sigma}(x)=-\tilde{R}_{n,\sigma}(-x)$.
\begin{equation}
 \tilde{\Psi}_{n,\sigma}(x)=e^{ik_{F}x}\tilde{R}_{n,\sigma}(x)-e^{-ik_{F}x} \tilde{R}_{n,\sigma}(-x)
\label{fermion}
\end{equation}
 We define a new chiral (right moving) Fermi field $\mathbf{\Omega}_{n,\sigma}(x)=\tilde{R}_{n,\sigma}(x)$ for $x>0$ and  $\mathbf{\Omega}_{n,\sigma}(x)=\tilde{L}_{n,\sigma}(-x)$ for $x<0$.
This implies that the chiral Fermionic field  $\mathbf{\Omega}_{n,\sigma}(x)$ obeys  $\mathbf{\Omega}_{n,\sigma}(L)=\mathbf{\Omega}_{n,\sigma}(-L)$.  $\mathbf{\Omega}_{n,\sigma}(x)$ is  periodic  in the domain $ -L<x<L$ ( the domain  $ 0<x<L$ has bee enlarged to $ -L<x<L$ ). Using the step function  $\theta[x]$ we write the representation of the  chiral field $\mathbf{\Omega}_{n,\sigma}(x)$.
\begin{equation}
\mathbf{\Omega}_{n,\sigma}(x)=\tilde{R}_{n,\sigma}(x) \theta[x]+ \tilde{L}_{n,\sigma}(-x) \theta[-x]
\label{symmetric}
\end{equation}
We find:
\begin{eqnarray}
&&H=H_{0}+H_{t},\hspace{0.05 in} H_{t}=H_{t,\sigma=\uparrow}+H_{t,\sigma=\downarrow} \nonumber\\&&
H_{0}= \sum_{\sigma=\uparrow,\downarrow}\sum_{n=1}^{N}\int_{-L}^{L}\,dx \mathbf{\Omega}^{\dagger}_{n,\sigma}(x)(-i\partial_{x})\mathbf{\Omega}_{n,\sigma}(x) ;\hspace{0.05 in} L\rightarrow \infty\nonumber\\&&
H_{t,\sigma=\uparrow}=t\sum_{n=1}^{N-1}\int_{0}^{L}\,dx\Big[ \tilde{R}_{n+1,\uparrow}(x)\tilde{R}_{n,\uparrow}(-x)+h.c.\Big] ;\hspace{0.05 in} L\rightarrow \infty\nonumber\\&&
H_{t,\sigma=\downarrow}=t\sum_{n=1}^{N-1}\int_{0}^{L}\,dx\Big[ \tilde{R}_{n+1,\downarrow}(-x)\tilde{R}_{n,\downarrow}(x)+h.c.\Big] ;\hspace{0.05 in} L\rightarrow \infty\nonumber\\&&
\end{eqnarray}
  The bulk is gaped and only four chiral  modes  remain gapless,  
\begin{equation}
\mathbf{\Omega}_{n=1,\uparrow}(x),\hspace{0.02in} \mathbf{\Omega}^{\dagger}_{n=1,\downarrow}(-x),\hspace{0.02 in} \mathbf{\Omega}_{n=N,\uparrow}(-x),\hspace{0.02 in} \mathbf{\Omega}^{\dagger}_{n=N,\downarrow}(x)
\label{edge}
\end{equation}
The chiral edge Hamiltonian is given by $H_{chiral ,n=1}=H_{left -edge}$, $H_{chiral ,n=N}=H_{right-edge}$: 
\begin{eqnarray}
&&H_{chiral}=H_{ left -edge}+H_{ right-edge}; \hspace{0.05 in} L\rightarrow \infty \nonumber\\&&
H_{ left- edge}= \int_{- \infty}^{\infty}\,dx\Big[ \mathbf{\Omega}^{\dagger}_{1,\uparrow}(x)(-i\partial_{x})\mathbf{\Omega}_{1,\uparrow}(x)+\mathbf{\Omega}^{\dagger}_{n=1,\downarrow}(x)(-i\partial_{x})\mathbf{\Omega}_{1,\downarrow}(x)\Big]\nonumber\\&&
H_{ right- edge}=\int_{-\infty}^{\infty}\,dx\Big[\mathbf{\Omega}^{\dagger}_{N,\downarrow}(x)(-i\partial_{x})\mathbf{\Omega}_{N,\downarrow}(x)  +\mathbf{\Omega}^{\dagger}_{N,\uparrow}(x)(-i\partial_{x})\mathbf{\Omega}_{N,\uparrow}(x)\Big]\nonumber\\&&
\end{eqnarray}
Using the proximity to a superconductor  with the  pairing field  $\Delta(x)$   we can gap out  the edges (the bulk states are gaped)  without breaking time reversal symmetry.  
\begin{eqnarray}
&&H=\int_{-\infty}^{\infty}\,dx\Big[\sum_{\sigma=\uparrow,\downarrow}\mathbf{\hat{\Omega}}^{\dagger}_{1,\sigma}(x)(-i\partial_{x})\mathbf{\hat{\Omega}}_{1,\sigma}(x)+   \sum_{\sigma=\uparrow,\downarrow}\mathbf{\hat{\Omega}}^{\dagger}_{N,\sigma}(x)(-i\partial_{x})\mathbf{\hat{\Omega}}_{N,\sigma}(x)
+\nonumber\\&&\Big(\Delta(x)e^{i\delta_{1}}\mathbf{\hat{\Omega}}^{\dagger}_{1,\uparrow}(x)\mathbf{\hat{\Omega}}^{\dagger}_{1,\downarrow}(x)+ \Delta(x)e^{i\delta_{N}}\mathbf{\hat{\Omega}}^{\dagger}_{N,\uparrow}(x)\mathbf{\hat{\Omega}}^{\dagger}_{N,\downarrow}(x)+H.C.\Big)\Big]
\nonumber\\&&
\end{eqnarray}
In the presence of a magnet which breaks reversal-symmetry the  spectrum  will also  be gaped out  \cite{Oreg}. 

\vspace{0.1 in}
\noindent
\textbf{2.2 The Fractional topological Insulator for  the filling factor  $\nu=\frac{k_{F}}{k_{so}}=\frac{1}{3}$}

\vspace{0.1 in}
\noindent
Next we will consider the model at the   filling factor   $\nu=\frac{k_{F}}{k_{so}}=\frac{1}{3}$.  We   use  composite Fermions in one dimensions  and  Bosonize the model around $3k_{F} $ (we mention that in one dimensions we can Bosonize around any odd number of Fermi momentum $k_{F}$).
In this section we will show how the method of composite Fermions works in one dimensions.

\noindent
  According to Eq.$(6)$ a composite Fermions is obtained  whenever    an even  number of  Jordan Wigner phases is attached to a Fermion. If  $\frac{1}{\sqrt{2\pi a}}e^{\pm i\pi \int_{-\infty}^{x}\,dx' n_{n,\sigma}(x')}e^{-i\sqrt{\pi}\varphi_{n,\sigma}(x)}$ describes an electrons, a composite fermions is obtained  by  modifying the Jordan Wigner phase to
 $\frac{1}{\sqrt{2\pi a}}e^{\pm i (2n+1)\pi \int_{-\infty}^{x}\,dx' n_{n,\sigma}(x')}e^{-i\sqrt{\pi}\varphi_{n,\sigma}(x)}$.
As a result one observes that the Bosonic representation for the composite fermions  with  $(2n+1)=3$  is obtained  for $\nu=\frac{k_{F}}{k_{so}}=\frac{1}{3}$. As a result the Bosonization   is invariant under the fermi momentum and filling factor mapping : $3k_{F}\rightarrow k_{F}$  and $\nu=\frac{1}{3}\rightarrow \nu=1$. Following the steps given in Eq.$(6)$ for  the filling factor $\nu=\frac{1}{3}$ we find:
\begin{eqnarray}
&&\tilde{\Psi}_{n,\sigma;c}(x)=\frac{1}{\sqrt{2\pi a}}F_{n,\sigma}e^{\pm i3\pi \int_{-\infty}^{x}\,dx' n_{n,\sigma}(x')}e^{-i\sqrt{\pi}\varphi_{n,\sigma}(x)}\nonumber\\&&
n_{n,\sigma}(x)=\bar{n}_{\sigma}+\frac{1}{\sqrt{\pi}}\partial_{x}\theta_{n,\sigma}(x), \pi\bar{n}_{n,\sigma}=k_{F},\partial_{x}\theta_{n,\sigma}(x)\equiv p_{n,\sigma}(x)\nonumber\\&&[\theta_{n,\sigma}(x),p_{n',\sigma'}(x')]_{-}=i\delta(x-x')\delta_{n,n'}\delta_{\sigma,\sigma'}\nonumber\\&&
\tilde{R}_{n,\sigma;c}(x)=\frac{1}{\sqrt{2\pi a}}F_{n,\sigma}e^{i\sqrt{\pi}(3\theta_{n,\sigma}(x)-\varphi_{n}(x)}\equiv\frac{1}{\sqrt{2\pi a}} F_{n,\sigma} e^{i\sqrt{4\pi}(2\theta^{R}_{n,\sigma}(x)+\theta^{L}_{n,\sigma}(x))}\nonumber\\&&
\tilde{L}_{n,\sigma;c}(x)=\frac{1}{\sqrt{2\pi a}}F_{n,\sigma}e^{-i\sqrt{\pi}(3\theta_{n,\sigma}(x)+\varphi_{n,\sigma}(x))}\equiv \frac{1}{\sqrt{2\pi a}} F_{n,\sigma}e^{-i\sqrt{4\pi}(2\theta^{L}_{n,\sigma}(x)+\theta^{R}_{n,\sigma}(x))}\nonumber\\&&
\tilde{\Psi}_{n,\sigma;c}(x)=e^{i3k_{F}x}\tilde{R}_{n,\sigma;c}(x)+e^{-i3k_{F}x} \tilde{L}_{n,\sigma;c}(x)\nonumber\\&&
\end{eqnarray}
Repeating the formulation given in Eqs.$(7-8)$ we have
\begin{eqnarray}
&& \tilde{\Psi}_{n,\sigma;c}(x)=e^{i3k_{F}x}\tilde{R}_{n,\sigma;c}(x)-e^{-i3k_{F}x} \tilde{R}_{n,\sigma; c}(-x)\nonumber\\&&
\mathbf{\Omega}_{n,\sigma;c}(x)=\tilde{R}_{n,\sigma;c}(x) \theta[x]+ \tilde{L}_{n,\sigma;c}(-x) \theta[-x]\nonumber\\&&
\end{eqnarray}
In the next step we use the relation $3k_{F}=k_{so}$
and obtain similar expressions to Eqs. $(8-12)$.
The bulk is gaped and only four chiral  modes  remain gapless  
\begin{equation}
\mathbf{\Omega}_{n=1,\uparrow;c}(x),\hspace{0.02in} \mathbf{\Omega}^{\dagger}_{n=1,\downarrow;c}(-x),\hspace{0.02 in} \mathbf{\Omega}_{n=N,\uparrow;c}(-x),\hspace{0.02 in} \mathbf{\Omega}^{\dagger}_{n=N,\downarrow;c}(x)
\label{edge}
\end{equation}
The chiral edge Hamiltonian is given by $H_{chiral ,n=1;c}=H_{left -edge;c}$, $H_{chiral ,n=N;c}=H_{right-edge;c}$: 
\begin{eqnarray}
&&H_{chiral;c}=H_{ left -edge;c}+H_{ right-edge;c}; \hspace{0.05 in} L\rightarrow \infty \nonumber\\&&
H_{ left- edge;c}= \int_{- \infty}^{\infty}\,dx\Big[ \mathbf{\Omega}^{\dagger}_{1,\uparrow;c}(x)(-i\partial_{x})\mathbf{\Omega}_{1,\uparrow;c}(x)+\mathbf{\Omega}^{\dagger}_{n=1,\downarrow;c}(x)(-i\partial_{x})\mathbf{\Omega}_{1,\downarrow;c}(x)\Big]\nonumber\\&&
H_{ right- edge}=\int_{-\infty}^{\infty}\,dx\Big[\mathbf{\Omega}^{\dagger}_{N,\downarrow}(x)(-i\partial_{x})\mathbf{\Omega}_{N,\downarrow;c}(x)  +\mathbf{\Omega}^{\dagger}_{N,\uparrow;c}(x)(-i\partial_{x})\mathbf{\Omega}_{N,\uparrow;c}(x)\Big]\nonumber\\&&
\end{eqnarray}
 Using  the relation imposed by the open boundary conditions with only  one  independent Bosonic field $\theta^{R}_{n,\sigma}(x)$ we have: 
\begin{equation}
\theta^{R}_{n,\sigma}(x)=\eta_{n,\sigma}(x),\hspace{0.05 in} \theta^{L}_{n,\sigma}(x)=\eta_{n,\sigma}(-x)
\label{ope}
\end{equation}
We build from  $\eta_{n,\sigma}(x)$ and $\eta_{n,\sigma}(-x)$ non-chiral  Bosonic fields $\Theta_{n,\sigma}(x)$ ,  $\Phi_{n,\sigma}(x)$:
\begin{equation}
\Theta_{n,\sigma}(x)=\eta_{n,\sigma}(x)+\eta_{n,\sigma}(-x),\hspace{0.01 in} \Phi_{n,\sigma}(x)=\eta_{n,\sigma}(-x)-\eta_{n,\sigma}(x)
\label{bos}
\end{equation}  
\begin{eqnarray}
&&
H_{ left- edge;c}= 
\int_{- \infty}^{\infty}\,dx\sum_{\sigma=\uparrow,\downarrow}\frac{v}{2}\Big[\kappa  (\partial_{x}\Phi_{n=1,\sigma}(x))^2+\frac{1}{\kappa} (\partial_{x}\Theta_{n=1,\sigma}(x))^2 
\Big]\nonumber\\&&
H_{ right- edge}=\int_{-\infty}^{\infty}\,dx\sum_{\sigma=\uparrow,\downarrow}\frac{v}{2}\Big[
\kappa  (\partial_{x}\Phi_{n=N,\sigma}(x))^2+\frac{1}{\kappa } (\partial_{x}\Theta_{n=N,\sigma}(x))^2 
\Big]\nonumber\\&&
v=3v_{0},\hspace{0.05 in} \kappa=\nu=\frac{k_{F}}{k_{so}}=\frac{1}{3}\nonumber\\&&
\end{eqnarray}
This shows that model is a Luttinger liquid with the  parameter $\kappa=\nu=\frac{1}{3}$. When the chains are in the proximity with a superconductor we add to the Luttinger liquid Hamiltonian in Eq.$(19)$ the pairing part given in Eq.$(12)$  (second line in Eq.$(12)$). As result    the  model  of the coupled chains in  proximity to a superconductor gives rise to a  $F.T.I.$. Following ref.\cite{Halperin} the F.T.I.  is identified  with the help of the  Josephson  periodicity which  measure  the degeneracy of the ground state.  

\vspace{0.2 in}

\noindent
\textbf{3.1.-The metallic disk in the presence of the   spin orbit interaction -a realization of a topological insulator}

\vspace{0.2 in}

\noindent
In this section we present our model.
it was shown that in strong magnetic we can use the limit of   large magnetic field  study the physics of electrons in strong magnetic fields   \cite{Iso}. we Using the analogy with the strong magnetic field we propose to study the spin -orbit interaction in the limit   $\frac{k^2_{so}}{m^*}\rightarrow \infty$.  As a result   a one  dimensional model  in a  confining potential  emerges. For    a parabolic potential $V(x,y)$  with the condition    $\frac{k^2_{so}}{m^*}\rightarrow \infty$ we find a constrained Hamiltonian, 

$h=\frac{\hbar^{2}}{2m^*}\Big[(-i\partial_{x}-\sigma_{z}k_{so} \frac{ y}{a})^2+ (-i\partial_{y})^2 \Big]-\mu+V(\vec{x},\vec{y})$ 

\noindent
 In the limit   $\frac{k^2_{so}}{m^*}\rightarrow \infty$  we obtain   $V(x,y=\sigma_{z}\frac{k_{x}}{k_{so}}a ) -\mu$.  The   two dimensional parabolic potential  $V(x,y)$  is  replaced by   a one dimensional model with a parabolic  potential. In  the second quantized formulation we find:  
\begin{equation}
H=\sum_{\sigma=\uparrow,\downarrow}\int\,dx \Psi^{\dagger}_{\sigma}(x)\Big[V(x,y=\sigma_{z}\frac{k_{x}}{k_{so}}a ) -\mu\Big]\Psi^{\dagger}_{\sigma}(x) 
\label{eq}
\end{equation}
Due to the constrained   $\frac{k^2_{so}}{m^*}\rightarrow \infty$ the   potential  $V(x,y)=\frac{g}{2}(x^2+y^2)$  is replaced by a one dimensional model.  The  coordinate  $y$ acts as the  momentum  $k_{x}$, and only $x$ remains the $x$ coordinate.
We find:
\begin{eqnarray}
&&H=\sum_{\sigma=\uparrow,\downarrow}\int\,dx \Psi^{\dagger}_{\sigma}(x)\Big[V(x,y=\sigma_{z}\frac{-i\partial_{x}}{k_{so}}a ) -\mu\Big]\Psi_{\sigma}(x)=\nonumber\\&&
\sum_{\sigma=\uparrow,\downarrow}\int\,dx \Psi^{\dagger}_{\sigma}(x)\Big[\frac{g a^2}{2 k^2_{so}}(-i\partial_{x})^2 +\frac{g }{2}x^{2}-\mu\Big]\Psi_{\sigma}(x)\nonumber\\&&
\end{eqnarray}
In the second line of Eq.$(21)$  we have used the constraint relation which emerges from the strong spin-orbit interaction $y=\frac{k_{x}}{k_{so}}a$,
This result is interpreted as a second class constrained \cite{Weinberg,Weir}
The one dimensional effective model given in Eq.$(21)$ with the  potential $\frac{g }{2}x^{2}$  allows to introduce a space dependent Fermi momentum, 
$k_{F}(x)=k_{so}\sqrt{\frac{2\mu}{ g a^2}}\Big(1-(\frac{x}{\hat{x}})^2\Big)^{\frac{1}{2}}$,
 where    $\hat{x}=\pm\sqrt{\frac{2 \mu}{g}}$ are  the  classically turning points.
We  can map this problem to the edge of the disk. We introduce  the  angular variables $\alpha(x)$ for the edge  ($\alpha$ is the angular variable for the edge which is a function of the original coordinate  $x$ ).
The mapping between the space dependent Fermi momentum and the angular variable  is given by the  function $\sin[\alpha(x)]$:

$\sqrt{1-(\frac{x}{\hat{x}})^{2})}=\sin[\alpha(x)]$  for    $ \pi \leq \alpha(x)\leq 0$, 

$-\sqrt{1-(\frac{x}{\hat{x}})^{2})}=\sin[\alpha(x)]$ for $ \pi \leq\alpha(x)\leq 2\pi $.

\noindent
 The turning points     $\hat{x}=\pm\sqrt{\frac{2 \mu}{g}}$ causes the  vanishing of the field $\Psi_{\sigma}(x)$ . For this reason  we must use open boundary conditions. As a result we     Bosonize $\Psi_{\sigma}(x)$  in terms of a single mover .
\begin{eqnarray}
&&\Psi_{\sigma}(x)=\frac{1}{\sqrt{2\pi a}}e^{\pm i\pi \int_{-\infty}^{x}\,dx' n_{\sigma}(x')}e^{-i\sqrt{\pi}\varphi_{\sigma}(x)}\nonumber\\&&
n_{\sigma}(x)=\bar{n}(x)+\frac{1}{\sqrt{\pi}}\partial_{x}\theta_{\sigma}(x), \pi\bar{n}(x)=k_{F}(x)\nonumber\\&&
\Psi_{\sigma}(x)=e^{i\int_{-\hat{x}}^{x}\,dx'k_{F}(x')}R_{\sigma}(x)-e^{-i\int_{-\hat{x}}^{x}\,dx'k_{F}(x')}R_{\sigma}(-x)\nonumber\\&&
\end{eqnarray}
    The Fermi  momentum  is a function of the chemical potential $\mu$ instead of two Fermi points $\pm k_{F}$ the   Fermi  momentum $ k_{F}(x)$ is  $x$ dependent. The vanishing points  $ k_{F}(x)=0 $ give rise to the effective edge for the disk.   $ k_{F}(x)$ is given by,

 $ k_{F}(x)=k_{so}\sqrt{\frac{2\mu}{ g a^2}}\sqrt{1-(\frac{x}{\hat{x}})^2}$ 

Due to the fact that the Fermi momentum is $x$ dependent  we  that Fermi velocity is also   space dependent,

 $v(x)=\sqrt{\frac{\mu g a^2}{2}}k_{so} \sqrt{1-(\frac{x}{\hat{x}})^2}$ 
 
In the next step we obtain the Bosonic  representation for the metallic disk.
\begin{eqnarray}
&&H=H^{(y>0)}+H^{(y<0)},\nonumber\\&&
H^{(y>0)}=\sum_{\sigma=\uparrow,\downarrow}\int_{-\hat{x}}^{\hat{x}}\,dx\Big[R^{\dagger}_{\sigma}(x;y>0)v(x)(-i\partial_{x})R_{\sigma}(x;y>0)-L^{\dagger}_{\sigma}(x;y>0)v(x)(-i\partial_{x})L_{\sigma}(x;y>0)\Big] \nonumber\\&&
H^{(y<0)}=\sum_{\sigma=\uparrow,\downarrow}\int_{-\hat{x}}^{\hat{x}}\,dx\Big[R^{\dagger}_{\sigma}(x;y<0)v(x)(-i\partial_{x})R_{\sigma}(x;y<0)-L^{\dagger}_{\sigma}(x;y<0)v(x)(-i\partial_{x})L_{\sigma}(x;y<0)\Big]\nonumber\\&&
\end{eqnarray}
$H^{(y>0)}$ represents the Hamiltonian for the upper half disk and $H^{(y<0)}$ is the Hamiltonian for the lower half.
Due to the turning points we have  the relations:
\begin{equation} 
 L_{\sigma}(x;y>0)=-R_{\sigma}(-x;y>0) ,\hspace{0.1 in} L_{\sigma}(x;y<0)=-R_{\sigma}(-x;y<0)
\label{map}
\end{equation}
Using the boundary conditions given in Eq.$(24)$ we obtain for Eq.$(23)$ the representation:
\begin{eqnarray}
&&H^{(y>0)}=\nonumber\\&&\sum_{\sigma=\uparrow,\downarrow}\int_{-\hat{x}}^{\hat{x}}\,dx\Big[R^{\dagger}_{\sigma}(x;y>0)v(x)(-i\partial_{x})R_{\sigma}(x;y>0)-R^{\dagger}_{\sigma}(-x;y>0)v(x)(-i\partial_{x})R_{\sigma}(-x;y>0)\Big]
\nonumber\\&&\equiv \sum_{\sigma=\uparrow,\downarrow}\int_{-\hat{x}}^{\hat{x}}\,dx\Big[R^{\dagger}_{\sigma}(x;y>0)2v(x)(-i\partial_{x})R_{\sigma}(x;y>0)\Big]\nonumber\\&&
H^{(y<0)}=
\nonumber\\&&\sum_{\sigma=\uparrow,\downarrow}\int_{-\hat{x}}^{\hat{x}}\,dx\Big[R^{\dagger}_{\sigma}(x;y<0)v(x)(-i\partial_{x})R_{\sigma}(x;y<0)-R^{\dagger}_{\sigma}(-x;y<0)v(x)(-i\partial_{x})R_{\sigma}(-x;y<0)\Big]\nonumber\\&&\equiv\sum_{\sigma=\uparrow,\downarrow}\int_{-\hat{x}}^{\hat{x}}\,dx\Big[R^{\dagger}_{\sigma}(x;y<0)2v(x)(-i\partial_{x})R_{\sigma}(x;y<0)\Big]\nonumber\\&&
\end{eqnarray}
Next we map the problem to the edge of the disk. 
We find from the mapping $x \rightarrow \alpha $ the relation $\frac{d\alpha(x)}{dx} =\frac{1}{|\sin[\alpha(x)]| \hat{x}}$.   The term  
$v(x)\partial_{x}$ is replaced by the derivative on the boundary of the disk $\partial_{\alpha}$. 
\begin{eqnarray}
&&
v(x)\partial_{x}=v(x)\frac{d\alpha(x)}{dx} \partial_{\alpha} \equiv\frac{k_{so} g a}{2}\partial_{\alpha} \nonumber\\&&
\int_{-\hat{x}}^{\hat{x}}\,dxf(x)=\int_{0}^{\pi}\,\frac{d\alpha}{\frac{d\alpha(x)}{dx}}f(\alpha(x)) =\hat{x}\int_{0}^{\pi}\,d\alpha\sin[\alpha(x)]f(\alpha(x))\approx \hat{x}\frac{2}{\pi}\int_{0}^{\pi}\,d\alpha f(\alpha)\nonumber\\&&
\end{eqnarray}
We express the Hamiltonian in Eq.$ (25)$   in terms of the chiral Fermions  on the boundary $\alpha(x)$  of the  disk.  We have the mapping  $\Big [R_{\uparrow}(x),R_{\downarrow}(x)\Big]^{T}\rightarrow \Big [R_{\uparrow}(\alpha),R_{\downarrow}(\alpha)\Big]^{T}$ and find:
\begin{eqnarray}
&&H=\frac{\sqrt{2}k_{so}a}{\pi}\sqrt{ \mu g } \oint\,d\alpha \Big[R_{\uparrow}^{\dagger} (\alpha) (-i\partial_{\alpha})R_{\uparrow} (\alpha)
+R_{\downarrow}^{\dagger} (\alpha) (-i\partial_{\alpha})R_{\downarrow} (\alpha)\Big]
\nonumber\\&&
\end{eqnarray}
Nest we consider the proximity effect of a superconductor with the pairing field   $ \Delta(\alpha)e^{i\delta}$. As a result of the pairing field a superconducting gap is open on the edges. As a results the Hamiltonian with the pairing field  $ \Delta(\alpha)e^{i\delta}$  gives rise to the   Bosonized form of the $T.I.$ Hamiltonian:
\begin{eqnarray}
&&H=\frac{\sqrt{2}k_{so}a}{\pi}\sqrt{ \mu g } \oint\,d\alpha \Big[R_{\uparrow}^{\dagger} (\alpha) (-i\partial_{\alpha})R_{\uparrow} (\alpha)
+R_{\downarrow}^{\dagger} (\alpha) (-i\partial_{\alpha})R_{\downarrow} (\alpha)\Big]
\nonumber\\&&
 +\sqrt{\frac{8\mu}{g\pi^2}} \oint\,d\alpha 
\Big[\Delta(\alpha)e^{i\delta}\Big( R_{\uparrow}^{\dagger} (\alpha) R_{\downarrow}^{\dagger} (\alpha)+ R_{\uparrow}^{\dagger} (-\alpha) R_{\downarrow}^{\dagger} (-\alpha)\Big)+H.C.\Big)\Big]\nonumber\\&&
\end{eqnarray}

\vspace{0.1 in}

\textbf{3.2. The metallic disk in the presence of the   spin orbit interaction-a  composite Fermion  formulation  for  a $F.T.I.$}
\vspace{0.1 in}

For particular densities the composite fermions construction introduced by  \cite{Jain}   can be used . In one dimensions the Jordan Wigner construction allows to obtain composite Fermions.

\noindent
Repeating the procedure of a space dependent Fermi momentum introduced in section $3.1$  we find that 
the turning points depends on the chemical potential,  $\frac{k_{F}(x)}{k_{so}}=\frac{1}{a}\sqrt{\frac{2\mu}{g}-x^2}$. By changing the chemical potential to  $\frac{\mu}{s^2}$, $s>1$ we obtain $\frac{k_{F}(x)}{k_{so}}=\frac{1}{a}\sqrt{\frac{2\mu}{s^2 g}-x^2}$. The turning points  decreases  to $\frac{\hat{x}}{s}=\sqrt{\frac{2\mu }{g s^2}}$.

The construction of the composite fermions  leaves the position of the turning point invariant. The Jordan Wigner construction is based on the fact that  both Jordan Wigner  representations  $\frac{1}{\sqrt{2\pi}}e^{\pm 3 i\pi \int_{-\infty}^{x}\,dx' n_{\sigma}(x')}$  and  $\frac{1}{\sqrt{2\pi}}e^{\pm  i\pi \int_{-\infty}^{x}\,dx' n_{\sigma}(x')}$  describe  a Fermion.  The first representation represents an interacting Fermion model with the filling factor $\frac{1}{3}$.  The second one represents a non interacting Fermion model with  the filling factor  $1$. For the chemical potential $\frac{\mu}{s^2}$  , the composite  Fermion with the  momentum $3k_{F}(x)$  will obey the relation  $3\frac{k_{F}(x)}{k_{so}}=\frac{3}{a}\sqrt{\frac{2\mu s^2}{g}-x^2}$. For 
$s=3$  we obtain $3\frac{k_{F}(x)}{k_{so}}=\frac{1}{a}\sqrt{\frac{2\mu}{g}-3^2x^2}$ giving the same turning point $\frac{\hat{x}}{3}=\sqrt{\frac{2\mu }{g 3^2}} $.

\noindent
For this case  we repeat the formulation given  in Eq.$(13)$. We replace $\theta_{\sigma}(x)$ and $\varphi_{\sigma}(x)$ with the chiral bosons 
$\theta^{L}_{\sigma}(x)$,$\theta^{R}_{\sigma}(x)$.  Due to the  boundary conditions at  the points $x=\pm \frac{\hat{x}}{3}$  we use the  relations  $\tilde{R}_{\sigma;c}(-x)=-\tilde{L}_{\sigma;c}(x)$.  We introduce $\theta^{R}_{\sigma}(x)\equiv\eta_{\sigma}(x)$ and $\theta^{L}_{\sigma}(x)\equiv\eta_{\sigma}(-x)$
\begin{eqnarray}
&&\tilde{\Psi}_{\sigma;c}(x)=\frac{1}{\sqrt{2\pi}}e^{\pm 3 i\pi \int_{-\infty}^{x}\,dx' n_{\sigma}(x')}e^{-i\sqrt{\pi}\varphi_{\sigma}(x)} \nonumber\\&&
n_{\sigma}(x)=\bar{n}(x)+\frac{1}{\sqrt{\pi}}\partial_{x}\theta_{\sigma}(x), \pi\bar{n}(x)=k_{F}(x)\nonumber\\&&
\theta_{\sigma}(x)=\theta^{R}_{\sigma}(x)+\theta^{L}_{\sigma}(x),\hspace{0.02 in}\varphi_{\sigma}(x)=\theta^{L}_{\sigma}(x)-\theta^{R}_{\sigma}(x)\nonumber\\&&
\tilde{R}_{\sigma;c}(x)=\frac{1}{\sqrt{2\pi}}e^{i\sqrt{\pi}(3\theta_{\sigma}(x)-\varphi_{\sigma}(x))}\equiv \frac{1}{\sqrt{2\pi}}e^{i\sqrt{4\pi}(2\theta^{R}_{\sigma}(x)+\theta^{L}_{\sigma}(x))}\equiv  \frac{1}{\sqrt{2\pi}}e^{i\sqrt{4\pi}(2\eta_{\sigma}(x)+\eta_{\sigma}(-x))}\nonumber\\&&
\tilde{L}_{\sigma;c}(x)=\frac{1}{\sqrt{2\pi}}e^{-i\sqrt{\pi}(3\theta_{\sigma}(x)+\varphi_{\sigma}(x))}\equiv \frac{1}{\sqrt{2\pi}} e^{-i\sqrt{4\pi}(2\theta^{L}_{\sigma}(x)+\theta^{R}_{\sigma}(x))}\equiv   \frac{1}{\sqrt{2\pi}}e^{-i\sqrt{4\pi}(2\eta_{\sigma}(-x)+\eta_{\sigma}(x))}\nonumber\\&&
\tilde{\Psi}_{\sigma;c}(x)=e^{i\int_{-\hat{x}}^{x}\,dx' 3k_{F}(x')}\tilde{R}_{\sigma;c}(x)-e^{-i\int_{-\hat{x}}^{x}\,dx'3k_{F}(x')}\tilde{R}_{\sigma;c}(-x)\nonumber\\&&
 \frac{3k_{F}(x)}{k_{so}}=\frac{1}{a}\sqrt{\frac{2\mu}{ g }}\sqrt{1-3^2x^2}\nonumber\\&&
\end{eqnarray}

We employ the mapping $x\rightarrow \alpha $ to the edge of the disk. When we a   superconductor is in the proximity   of  the disk the paring field $\Delta(\alpha)e^{i\delta}$  will generate a gap 
\begin{eqnarray}
&&H=\frac{\sqrt{2}k_{so}a}{\pi}\sqrt{ \mu g } \oint\,d\alpha \Big[\tilde{R}_{\uparrow}^{\dagger} (\alpha) (-i\partial_{\alpha})\tilde{R}_{\uparrow} (\alpha)
+\tilde{R}_{\downarrow}^{\dagger} (\alpha) (-i\partial_{\alpha})\tilde{R}_{\downarrow} (\alpha)\Big]+
\nonumber\\&& \sqrt{\frac{8\mu}{9g\pi^2}} \oint\,d\alpha 
\Big[\Delta(\alpha)e^{i\delta}\Big( R_{\uparrow}^{\dagger} (\alpha) R_{\downarrow}^{\dagger} (\alpha)+ R_{\uparrow}^{\dagger} (-\alpha) R_{\downarrow}^{\dagger} (-\alpha)\Big)+H.C.\Big)\Big]\nonumber\\&&
\end{eqnarray}
We introduce the fields:
\begin{eqnarray}
&&\Theta_{\sigma}(x)=\eta_{\sigma}(x)+\eta_{\sigma}(-x),\hspace{0.1 in} \Phi_{\sigma}(x)= \eta_{\sigma}(-x)-\eta_{\sigma}(x))\nonumber\\&&
 \Theta_{c}(x)= \frac{\Theta_{\uparrow}(x)+ \Theta_{\downarrow}(x)}{\sqrt{2}},\hspace{0.1 in}\Phi_{c}(x)=\frac{ \Phi_{\uparrow}(x)+ \Phi_{\downarrow}(x)}{\sqrt{2}}\nonumber\\&&
\end{eqnarray}
 $\partial_{x}\Theta_{c}(x)$ measures the charge density which is conjugate to $\Phi_{c}(x)$.
We map the Bosonic fields $\Theta_{c}(x)$ and $\Phi_{c}(x)$ to the edge of the disk; $\Theta_{c}(x)\rightarrow \Theta_{c}(\alpha)$, $ \Phi_{c}(x)\rightarrow \Phi_{c}(\alpha)$.
The Bosonic form of   the Hamiltonian in Ed.$(29)$ reveals the Luttinger liquid  structures with  the interacting  parameter , $\kappa=\nu=\frac{k_{F}}{k_{so}}=\frac{1}{3}$. As a result the charge sector represents  an $F.T.I.$.
\begin{eqnarray}
&&H=\frac{3\sqrt{2}k_{so}a}{\pi}\sqrt{\mu g} \oint\,d\alpha
\frac{v}{2}\Big[
\kappa  (\partial_{\alpha}\Phi_{c}(\alpha))^2+\frac{1}{\kappa} (\partial_{\alpha}\Theta_{c}(\alpha))^2 
\Big]+\nonumber\\&&\sqrt{\frac{128\mu}{9g\pi^2}} \oint\,d\alpha 
|\Delta(\alpha)| \cos[3\sqrt{2\pi}\Theta_{c}(\alpha)]\cos[\sqrt{2\pi} \Phi_{c}(\alpha) +\delta] \nonumber\\&&
 v=\frac{3\sqrt{2}k_{so}a}{\pi}\sqrt{ \mu g },\hspace{0.1in} \kappa=\nu=\frac{k_{F}}{k_{so}}=\frac{1}{3} \nonumber\\&&
\end{eqnarray}
Comparing the results in Eq.$(31)$  with the one given in Eq.$(27)$ we notice that $\kappa=1$ and the pairing operator is replaced by the symmetric form , $\cos[\sqrt{2\pi}\Theta_{c}(\alpha)]\cos[\sqrt{2\pi} \Phi_{c}(\alpha) +\delta]$.
As a result the Josephson current will be different for  the two cases.
The use of the zero mode operators given in ref.\cite{david,Boyanovsky}   can reveal the Josephson periodicity of the degenerate  ground state. 

When the superconductor is replaced by a  magnetic system   a gap  on the  edge  of the disk via spin-flipping backscattering will appear. In this case the Josephson charge current   will be replaced by a Josephson spin current \cite{Halperin}.

\noindent
 The experimental verification is done by measuring the Josephson  current between the metallic disk and the  superconductor which will show  different results for the  T.I. and the   F.T.I.
 
The experimental question is how to drive the disk to be either a T.I. or F.T.I. Our results show that for the two cases we have different turning points,  $\sqrt{\frac{2\mu}{g}}$ for a T.I. and $\sqrt{\frac{2\mu}{9g}}$ for a F.T.I..
The physical radius of the disk $R$ determines what state can be obtained. When the radius $R$ obeys $R> \sqrt{\frac{2\mu}{g}}$ the T.I. and the F.T.I. are  possible.  We will have a coherent or a mixture of the two phases. In order observe a single  phase we have to chose the radius to satisfy   $\sqrt{\frac{2\mu}{9g}}\leq R\leq \sqrt{\frac{2\mu}{g}}$. For this case the phase with $\nu=1$ is not possible ( the ring radius is shorter then the turning point ), from the other-hand  the F.T.I. with  $\nu=\frac{1}{3}$ is  possible to observe. 

\vspace{0.2 in}

\noindent

\textbf{4. Conclusions}

\vspace{0.2 in}

In the first of this paper we have presented the Bosonization for the model considered   in ref.\cite{Halperin}. We have found that is essential to use open boundary conditions.  This results were obtained using chiral Bosonization. The Fractional case has been obtained with the help of the Jordan Wigner transformation  for composite Fermions.

In the second part we   we propose a new model for a Fractional Topological Insulators.  We consider  a  metallic disk, and take advantage of the strong spin orbit interaction in the presence of a parabolic potential. We map the problem to an one dimensional model with a  harmonic potential.  On the edge of the disk we find  a  chiral  fermion model which in the proximity to  a superconductor gives rise to a Fractional Topological Insulator  when the radius of the disk is tuned to be larger then the fractional turning point. 

The mapping to the one dimensions allows to show that the  Fractional Topological Insulator emerges as  an effective Luttinger liquid model      for the filling factor $\nu=\frac{1}{3}$ .

A possible experimental  realization of the model is suggested,  based on tunning of the chemical potential and  the radius of the disk.

\end{document}